\documentclass[preprint2]{aastex6}

\usepackage{graphicx}

\shorttitle{CGPS 1420 MHz Compact Source Catalog}
\shortauthors{Taylor et.\ al.}

\begin{document}

\title{A 1420\,MHz Catalog of Compact Sources in the Northern Galactic Plane}

\author{A.\ R.\ Taylor}
\affil{Inter-University Institute for Data Intensive Astronomy, and \\
Department of Astronomy, University of Cape Town \\
Department of Physics, University of the Western Cape \\
}
\author{ D.\ A.\ Leahy, T. Wenwu, C. Sunstrum}
\affil{Department of Physics and Astronomy, University of Calgary}

\author{ R.\ Kothes, T.\ L.\ Landecker, R.\ R Ransom, L.\ A. Higgs} 
\affil{Dominion Radio Astrophysical Observatory, Herzberg Programs in Astronomy
and Astrophysics \\ National Research Council of Canada}

\begin{abstract}
We present a catalog of compact sources of radio emission at 1420 MHz in the
northern Galactic plane from  the Canadian Galactic Plane Survey.  The
catalog contains 72,787 compact sources with angular size less than 3$'$
within the Galactic longitude range $52^{\circ} < \ell < 192^{\circ}$
down to a 5$\sigma$  detection level of $\sim$1.2 mJy.  Linear polarization
properties are included for 12,368 sources with signals greater than
4$\sigma_{QU}$ in the CGPS Stokes $Q$ and $U$ images at the position of the
total intensity peak.  We compare CGPS flux densities with cataloged flux
densities in the Northern VLA Sky Survey catalog for 10,897  isolated
unresolved sources with CGPS flux density greater than 4 mJy to search for
sources that show variable flux density on time scales of several years. We
identify 146 candidate variables that exhibit high fractional variations between
the two surveys.  In addition we identify 13 candidate transient sources that
have CGPS flux density above 10\,mJy but are not detected in the NVSS.
\end{abstract}


\section{Introduction}
\label{intro}

The Canadian Galactic Plane Survey (CGPS) mapped the principal constituents of
the interstellar medium (ISM)  with high angular resolution, inspired by the
view that understanding the ISM requires knowledge of all its constituents.  The
Synthesis Telescope at the Dominion Radio Astrophysical Observatory (the DRAO
ST) provided arcminute-resolution  images of the radio continuum and
atomic-hydrogen line emission. The CGPS is presented in
\citet{2003AJ....125.3145T}, where details are also given of companion surveys, which
imaged the molecular and dust constituents of the ISM. The DRAO ST is described
in \citet{2000A&AS..145..509L}. Polarization imaging techniques and images of
the diffuse Galactic polarized emission at 1420 MHz are presented in
\citet{2010A&A...520A..80L}.

The CGPS DRAO radio continuum observations provide images of Stokes $I$, $Q$,
and $U$ in four 7.5-MHz sub-bands spanning 35 MHz, centred on 1420 MHz.
The observations were carried out in three phases beginning in 1995 and  ending
in 2009. The sky coverage of each phase and observing dates are listed in
Table~\ref{coverage}.  The Galactic plane was covered with a width in Galactic
latitude of $9^{\circ}$, centred at  $b = 1^{\circ}$ to accommodate the warp of
the Galactic disk. The longitude coverage was constrained by the southern
declination limit of $\sim$20$^{\circ}$, the range that could be effectively
imaged by a linear East-West synthesis telescope array. The Phase II
observations included an extension to higher latitudes ($b = 17.5^{\circ}$) over
a restricted range of longitude. 

\begin{deluxetable*}{lccc}
\tablecaption{ Galactic Coordinate Coverage of the CGPS Observations 
\label{coverage} }
\tablehead{
\colhead{Phase} & \colhead{Galactic Latitude} & \colhead{Galactic Longitude}  & \colhead{Dates} \\
& \colhead{(degrees)} & \colhead{(degrees)} }
\startdata
Phase I   &  $74 < \ell < 147$  &  $-3.6 < b < 5.6$  & Apr 1995 -- Jun 2000 \\
Phase II  &  $64 < \ell < 74$     &  $-3.6 < b < 5.6$ & Feb 2000 -- Apr 2004 \\
               &  $147 < \ell < 175$ &  $-3.6 < b < 5.6$   \\
               & $100 < \ell < 117$  &  $\quad\ 5.6 < b < 17.5$ \\
Phase III & $52 < \ell < 64 $ & $-3.6 < b < 5.6$ & Nov 2003 -- Feb 2009 \\
              &  $175 < \ell < 192$ &   $-3.6< b < 5.6$ \\
\enddata
\end{deluxetable*}

At radio wavelengths the plane of the Galaxy is characterized by strong and
highly-structured extended emission from the ionized and relativistic plasma of
the ISM. The presence of such complex emission structure on all angular scales
presents challenges for interferometric observations.   To minimize imaging
artefacts the CGPS observations fully sample the two-dimensional spatial
frequency distribution of the sky from zero spacing up to the maximum baseline
of 617 m, and direction-dependent self-calibration is used to remove the effects
of sources outside of the primary beams of the antennas. The CGPS 1420-MHz
images are thus noise-limited images largely free of imaging artefacts over most
of the survey area.    The  band-integrated images have a $1\sigma$ RMS noise as
low as 0.18 mJy. These data therefore  provide an opportunity to create a
sensitive catalog of radio continuum sources  over a large area of the northern
Galactic plane, including both total intensity and linear polarization.     

In this paper we present the CGPS 1420-MHz compact source catalog covering 1464
square degrees and spanning a range of $140^{\circ}$ of Galactic longitude
between $52^{\circ} < \ell < 192^{\circ}$. The catalog is presented and described in
Section~\ref{cat}, including flux densities in Stokes $I$ and polarized intensities.
We compare the CGPS compact source catalog with the Northern
VLA Sky Survey (NVSS) catalog \citep{1998AJ....115.1693C} in Section~\ref{nvss}.
Since the two surveys have comparable resolution and sensitivity, but were separated
in time, we are able to carry out a simple analysis for variability.  We identify a number of variable sources
and transient candidates in 
(Section~\ref{vari}).

\section{The Compact Sources Catalog}
\label{cat}

\subsection{The Source List} 
\label{list}

The CGPS DRAO image products are 5.12$^{\circ}$$\times$5.12$^{\circ}$ images
mosaicked from overlapping Synthesis Telescope fields.   The images are
available through the Canadian Astronomy Data Centre 
{\footnote{{\tt{www.cadc-ccda.hia-iha.nrc-cnrc.gc.ca/en/cgps}}}}.    The
complete survey is comprised of 85 such mosaics. We constructed the radio source
catalog from these images using an automated source  finding algorithm, FINDSRC,
developed as part of the DRAO export software  package
\citep{1997ASPC..125...58H}.   FINDSRC uses a wavelet filter convolution to
identify compact source candidates in the image.  This convolution effectively
filters out emission structures larger than a few beam sizes.   
 A second DRAO software program, FLUXFIT, reads in the candidates identified in the filtered 
image by FINDSRC and fits the signals in the original, unfiltered
image with a two-dimensional Gaussian function plus a ``twisted'' plane
fit to the local background level.   
To avoid the potential problem of nearby sources within the fitting region, if a nearby
sources is detected, FINDSRC expands the area of the region to include that
source and carries out a simultaneous multiple Gaussian fit to all sources 
in the region.
The output of FINDSRC are fit parameters and errors for the peak and integrated flux
density of the source above the local background, as well as the major and minor
axes and the major axis position angle of the best-fit Gaussian approximation to
the source response function. 
For each source an RMS background level was
calculated from the variance in the image around the twisted plane background
fit.  We retained a source in the catalog if the peak amplitude of the best-fit
Gaussian is greater than five times the background RMS.

Following the automated source extraction, an annotation file marking the
position of each detected source was overlaid on the mosaic images and each
image was examined visually to identify spurious detections arising, for example, from
knots on filaments of highly structured diffuse emission in the Galactic Plane. 
On a few occasions a source was missed by the
automated detection algorithm.  In this case Gaussian fits were run ``manually".
The final source list contains 72,787 sources down to a minimum Stokes $I$ 
peak intensity of $\sim$1 mJy~beam$^{-1}$.   

Because the DRAO ST is an East-West aperture synthesis array, the  major axis of
the synthesized beam has position angle of zero (oriented North-South) and its
dimension is a function of declination.  The angular resolution of the ST
observations is given by
\begin{equation}
{\rm b_{major} \times b_{minor}} = 49''{\rm cosec}\delta \times 49''
\label{eqn:beam}
\end{equation}
where ${\rm b_{major}}$ and ${\rm b_{minor}}$ are respectively in the
Declination and Right Ascension directions. A Gaussian synthesized beam is
deconvolved from the fitted two-dimensional Gaussian to estimate the
observed angular size of each source.  A source is considered unresolved in
either the major or minor axis dimension if the deconvolved angular size is less
than 3 times the 1$\sigma$ error on the solid angle.  
Unresolved dimensions are listed in the catalog with zero dimension.

A measure of the polarized intensity for each source was obtained from intensity
of the Stokes $Q$ and $U$  images at the Stokes $I$ position of each source
after removing a mean background level determined in a 25$\times$25 pixel region
around each source. A provisional polarization detection was flagged if either
the $Q$ or $U$ signal was greater than four  times the local RMS, $\sigma_{Q}$
and $\sigma_{U}$, in the region surrounding the source.     For provisional
detections an estimate, $\hat p_o$,  of the intrinsic polarized intensity,
$p_o$, was calculated as
\begin{equation} \hat p_o = \sqrt{p^2 - \sigma_{\rm QU}^2}, 
\end{equation} 
which is good estimator of the polarized intensity when $p/\sigma > 4 $
\citep{1985A&A...142..100S}.  A source was retained as a polarized detection if
in addition $\hat p_o > \epsilon_p S_p$, where $\epsilon_p$ is the instrumental
polarization error, taken to be 0.3\% \citep{2003AJ....125.3145T}, and $S_p$ is
the peak Stokes $I$ intensity. The error on the polarized intensity is given by 
\begin{equation}
\sigma_{\hat p_o} = \sqrt{2 \sigma_{\rm QU}^2 + (\epsilon_p S_p)^2}.
\end{equation}   

We detect polarized signals from 12,368 sources. The
polarized intensity should be treated with caution for resolved sources.  The
polarized intensity structure for extended sources is not necessarily the same
as that of the total intensity.  In such cases the polarized intensity at the
position of the total intensity peak may not represent the true peak of $Q$ and
$U$.  Those interested in the polarized emission from extended sources are
advised to examine the CGPS images directly.

Table~\ref{catalog} lists the first thirty entries of the CGPS 1420 MHz
catalog.  The columns contain: 
(1) the source name CGPS JHHMMSS+DDMMSS, where HHMMSS are the hours and minutes 
and seconds of time for Right Ascension, and DDMMSS are the degrees, minutes and seconds of arc of
Declination (both in J2000),
(2) the J2000 Right Ascension (degrees) and error (seconds),
(3) the J2000 Declination (degrees) and error (arcseconds), 
(4) the integrated Stokes $I$ flux density and error (mJy),
(5) the peak Stokes $I$ intensity and error (mJy/beam), 
(6) the average rms around the position of the source in the $Q$ and $U$ images (mJy),
(7) the bias corrected polarized intensity in the case of detected polarized 
emission (mJy/beam),
(8) the polarization position angle (degrees),
(9 and 10) the deconvolved dimensions of the major and minor axes of the Stokes
$I$ emission with errors (arc seconds), and 
(11) the position angle of the source major axis (degrees) with respect to north.
The deconvolved source dimensions are listed as zero if the source is
classed as unresolved.  The beam dimensions at the location of the source can 
be calculated using Equation~\ref{eqn:beam}.  All errors are $1\sigma$.  
 
 The complete version of Table~\ref{catalog} is available in machine-readable
 format through the CGPS data repository at the Canadian Astronomy Data Centre.

\begin{deluxetable*}{cccrrrrcrrrr}
\tabletypesize{\tiny}
\tablewidth{\textwidth}
\tablecaption{ CGPS 1420 MHz Source Catalog\label{catalog} }
\tablehead{
\colhead{Name}& \colhead{RA(J2000)} & \colhead{DEC(J2000)} & 
\colhead{S$_{\rm I}$} & \colhead{S$_{\rm P}$} & \colhead{$\sigma_{QU}$} &
\colhead{P} & \colhead{Pol PA} & \colhead{$\theta_{\rm maj}$} & \colhead{$\theta_{\rm min}$} & 
\colhead{PA} \\
&&& \colhead{(mJy)} &  \colhead{(mJy/bm)} & \colhead{(mJy)} & \colhead{(mJy/bm)} &
\colhead{($\circ$)} & \multicolumn{2}{c}{(arc-seconds)} & \colhead{($\circ$)} 
}
\startdata
CGPS J000001+610821 &   0.00508 $\pm$  0.47 & 61.13939 $\pm$  4.6 &      2.6 $\pm$   0.31 &      2.6 $\pm$   0.31 &   0.27 &      0.0 $\pm$   0.00&    0 &    0.0 $\pm$  0.0 &    0.0 $\pm$  0.0 &    0 \\ 
CGPS J000002+670758 &   0.00887 $\pm$  0.30 & 67.13281 $\pm$  0.7 &     37.1 $\pm$   1.24 &     30.7 $\pm$   1.12 &   0.31 &      2.2 $\pm$   0.32&   69 &   29.3 $\pm$  0.8 &   15.3 $\pm$  1.4 &   76 \\ 
CGPS J000002+605312 &   0.00996 $\pm$  0.66 & 60.88678 $\pm$ 13.4 &      1.2 $\pm$   0.27 &      1.2 $\pm$   0.27 &   0.34 &      0.0 $\pm$   0.00&    0 &    0.0 $\pm$  0.0 &    0.0 $\pm$  0.0 &    0 \\ 
CGPS J000002+620952 &   0.01162 $\pm$  1.07 & 62.16464 $\pm$  6.0 &      1.2 $\pm$   0.25 &      1.2 $\pm$   0.25 &   0.30 &      0.0 $\pm$   0.00&    0 &    0.0 $\pm$  0.0 &    0.0 $\pm$  0.0 &    0 \\ 
CGPS J000003+625231 &   0.01650 $\pm$  0.40 & 62.87531 $\pm$  2.2 &      8.1 $\pm$   0.47 &      8.1 $\pm$   0.47 &   0.30 &      1.6 $\pm$   0.29&  -72 &    0.0 $\pm$  0.0 &    0.0 $\pm$  0.0 &    0 \\ 
CGPS J000005+670622 &   0.02346 $\pm$  0.57 & 67.10622 $\pm$  6.7 &      2.0 $\pm$   0.28 &      2.0 $\pm$   0.28 &   0.32 &      0.0 $\pm$   0.00&    0 &    0.0 $\pm$  0.0 &    0.0 $\pm$  0.0 &    0 \\ 
CGPS J000006+592611 &   0.02567 $\pm$  0.48 & 59.43644 $\pm$  4.6 &      2.6 $\pm$   0.32 &      2.6 $\pm$   0.32 &   0.36 &      0.0 $\pm$   0.00&    0 &    0.0 $\pm$  0.0 &    0.0 $\pm$  0.0 &    0 \\ 
CGPS J000007+624215 &   0.03021 $\pm$  0.53 & 62.70439 $\pm$  3.9 &      3.2 $\pm$   0.31 &      3.2 $\pm$   0.31 &   0.34 &      0.0 $\pm$   0.00&    0 &    0.0 $\pm$  0.0 &    0.0 $\pm$  0.0 &    0 \\ 
CGPS J000007+660841 &   0.03133 $\pm$  0.31 & 66.14472 $\pm$  0.9 &     10.7 $\pm$   0.49 &      9.9 $\pm$   0.40 &   0.21 &      1.0 $\pm$   0.22&   26 &   19.8 $\pm$  2.0 &    0.0 $\pm$  0.0 &   66 \\ 
CGPS J000015+624330 &   0.06450 $\pm$  0.32 & 62.72511 $\pm$  1.3 &     11.7 $\pm$   0.47 &     11.7 $\pm$   0.47 &   0.30 &      0.0 $\pm$   0.00&    0 &    0.0 $\pm$  0.0 &    0.0 $\pm$  0.0 &    0 \\ 
CGPS J000015+611227 &   0.06525 $\pm$  0.57 & 61.20769 $\pm$  4.1 &      2.9 $\pm$   0.28 &      2.9 $\pm$   0.28 &   0.30 &      0.0 $\pm$   0.00&    0 &    0.0 $\pm$  0.0 &    0.0 $\pm$  0.0 &    0 \\ 
CGPS J000015+625015 &   0.06550 $\pm$  0.30 & 62.83761 $\pm$  0.7 &     78.0 $\pm$   2.59 &     48.9 $\pm$   2.31 &   0.37 &      0.0 $\pm$   0.00&    0 &   57.0 $\pm$  0.8 &   19.5 $\pm$  1.3 &  128 \\ 
CGPS J000016+621817 &   0.06904 $\pm$  0.66 & 62.30492 $\pm$  7.2 &      3.1 $\pm$   0.49 &      3.1 $\pm$   0.49 &   0.35 &      0.0 $\pm$   0.00&    0 &    0.0 $\pm$  0.0 &    0.0 $\pm$  0.0 &    0 \\ 
CGPS J000017+631526 &   0.07125 $\pm$  0.53 & 63.25731 $\pm$  4.3 &      3.2 $\pm$   0.52 &      2.3 $\pm$   0.28 &   0.25 &      0.0 $\pm$   0.00&    0 &   39.0 $\pm$  8.4 &   23.7 $\pm$  0.2 &  144 \\ 
CGPS J000017+605635 &   0.07271 $\pm$  0.30 & 60.94317 $\pm$  0.7 &     29.8 $\pm$   1.03 &     26.7 $\pm$   0.91 &   0.30 &      0.0 $\pm$   0.00&    0 &   21.3 $\pm$  1.1 &   12.2 $\pm$  2.1 &   97 \\ 
CGPS J000018+622221 &   0.07662 $\pm$  0.32 & 62.37258 $\pm$  1.3 &     11.2 $\pm$   0.47 &     11.2 $\pm$   0.47 &   0.29 &      0.0 $\pm$   0.00&    0 &    0.0 $\pm$  0.0 &    0.0 $\pm$  0.0 &    0 \\ 
CGPS J000018+605251 &   0.07683 $\pm$  0.35 & 60.88100 $\pm$  2.5 &      4.5 $\pm$   0.31 &      4.5 $\pm$   0.31 &   0.32 &      0.0 $\pm$   0.00&    0 &    0.0 $\pm$  0.0 &    0.0 $\pm$  0.0 &    0 \\ 
CGPS J000019+630951 &   0.08088 $\pm$  0.30 & 63.16419 $\pm$  0.7 &     21.2 $\pm$   0.78 &     19.7 $\pm$   0.67 &   0.27 &      0.0 $\pm$   0.00&    0 &   19.3 $\pm$  1.6 &    8.5 $\pm$  2.8 &  168 \\ 
CGPS J000020+611404 &   0.08608 $\pm$  0.59 & 61.23453 $\pm$  2.9 &      3.1 $\pm$   0.28 &      3.1 $\pm$   0.28 &   0.23 &      0.0 $\pm$   0.00&    0 &    0.0 $\pm$  0.0 &    0.0 $\pm$  0.0 &    0 \\ 
CGPS J000021+610428 &   0.08862 $\pm$  0.49 & 61.07444 $\pm$  6.3 &      1.9 $\pm$   0.27 &      1.9 $\pm$   0.27 &   0.30 &      0.0 $\pm$   0.00&    0 &    0.0 $\pm$  0.0 &    0.0 $\pm$  0.0 &    0 \\ 
CGPS J000021+661831 &   0.08908 $\pm$  0.54 & 66.30878 $\pm$  8.2 &      2.2 $\pm$   0.51 &      1.2 $\pm$   0.23 &   0.20 &      0.0 $\pm$   0.00&    0 &   79.1 $\pm$  0.7 &    0.0 $\pm$  0.0 &   70 \\ 
CGPS J000023+642554 &   0.09617 $\pm$  0.31 & 64.43186 $\pm$  0.9 &     19.6 $\pm$   0.82 &     16.1 $\pm$   0.64 &   0.31 &      0.0 $\pm$   0.00&    0 &   28.0 $\pm$  1.7 &   20.0 $\pm$  1.9 &   23 \\ 
CGPS J000024+621506 &   0.10054 $\pm$  0.30 & 62.25178 $\pm$  0.6 &    336.8 $\pm$   9.91 &    277.4 $\pm$   9.84 &   0.35 &     10.7 $\pm$   0.90&   65 &   27.2 $\pm$  0.3 &   20.6 $\pm$  0.4 &   64 \\ 
CGPS J000025+641845 &   0.10438 $\pm$  0.43 & 64.31256 $\pm$  2.9 &      4.7 $\pm$   0.31 &      4.7 $\pm$   0.31 &   0.33 &      0.0 $\pm$   0.00&    0 &    0.0 $\pm$  0.0 &    0.0 $\pm$  0.0 &    0 \\ 
CGPS J000026+641327 &   0.11067 $\pm$  0.65 & 64.22419 $\pm$  6.3 &      1.6 $\pm$   0.29 &      1.6 $\pm$   0.29 &   0.30 &      0.0 $\pm$   0.00&    0 &    0.0 $\pm$  0.0 &    0.0 $\pm$  0.0 &    0 \\ 
CGPS J000027+632556 &   0.11287 $\pm$  0.43 & 63.43236 $\pm$  3.4 &      2.8 $\pm$   0.27 &      2.8 $\pm$   0.27 &   0.33 &      0.0 $\pm$   0.00&    0 &    0.0 $\pm$  0.0 &    0.0 $\pm$  0.0 &    0 \\ 
CGPS J000027+595049 &   0.11488 $\pm$  0.32 & 59.84711 $\pm$  1.3 &      6.9 $\pm$   0.32 &      6.9 $\pm$   0.32 &   0.43 &      0.0 $\pm$   0.00&    0 &    0.0 $\pm$  0.0 &    0.0 $\pm$  0.0 &    0 \\ 
CGPS J000028+661149 &   0.11937 $\pm$  0.32 & 66.19694 $\pm$  1.1 &      7.8 $\pm$   0.45 &      6.6 $\pm$   0.32 &   0.23 &      0.0 $\pm$   0.00&    0 &   23.3 $\pm$  2.3 &   19.4 $\pm$  2.9 &   66 \\ 
CGPS J000029+595722 &   0.12246 $\pm$  0.33 & 59.95636 $\pm$  1.8 &      5.8 $\pm$   0.32 &      5.8 $\pm$   0.32 &   0.42 &      0.0 $\pm$   0.00&    0 &    0.0 $\pm$  0.0 &    0.0 $\pm$  0.0 &    0 \\ 
CGPS J000029+593426 &   0.12429 $\pm$  0.30 & 59.57394 $\pm$  0.7 &     28.8 $\pm$   1.01 &     24.1 $\pm$   0.88 &   0.29 &      0.0 $\pm$   0.00&    0 &   25.9 $\pm$  1.0 &   20.8 $\pm$  1.3 &  125 \\ 
..... \\
\enddata
\tablecomments{Table~\ref{catalog} is published in its entirety in the machine-readable format.  A portion is shown here for 
guidance regarding its form and content.}
\end{deluxetable*}

\section{Comparison to the NVSS}
\label{nvss}
\subsection{Position and Flux Densities}
\label{posflux}

\begin{figure}[tbh]
\begin{center}
\includegraphics[width=1.05\columnwidth]{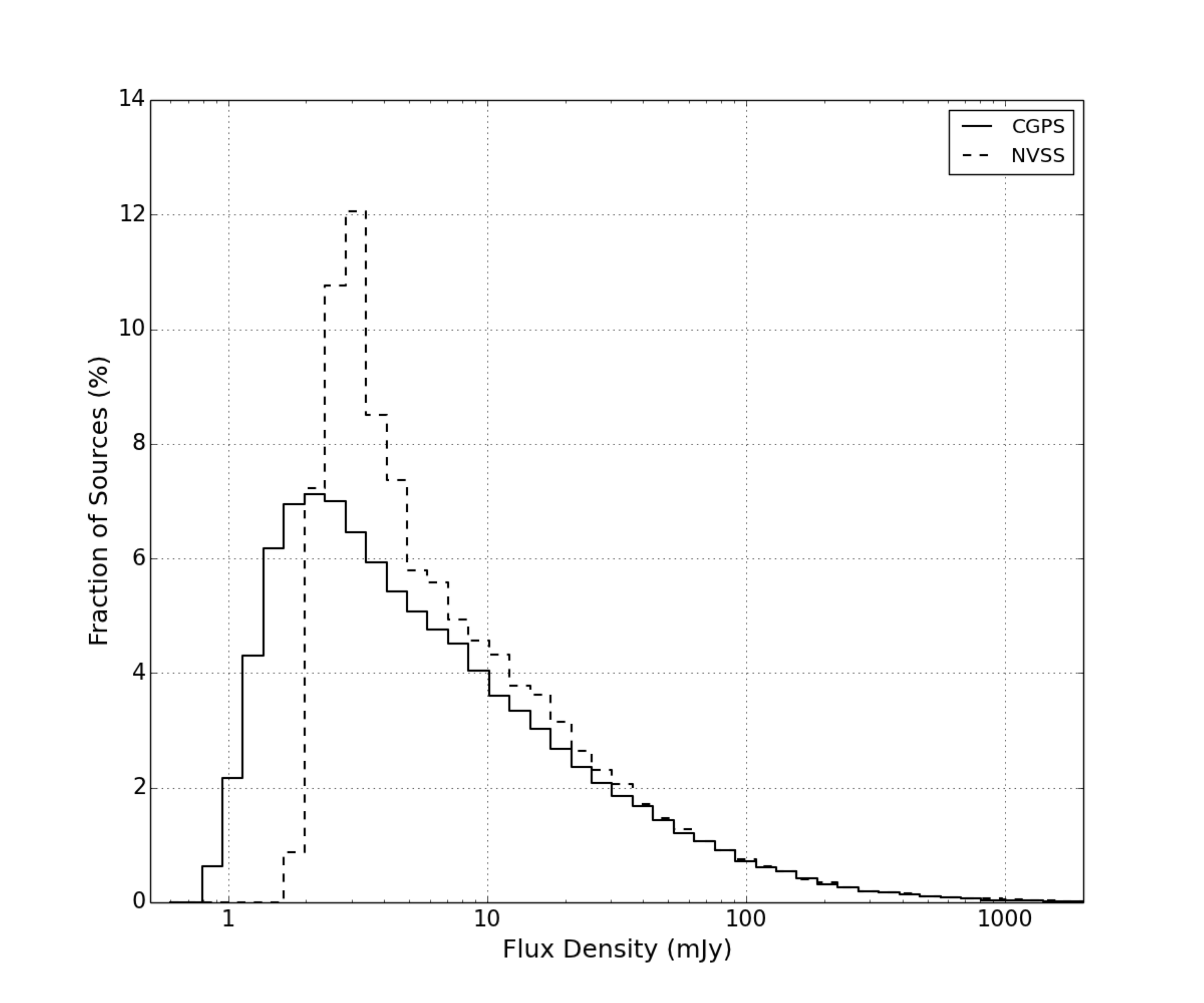}
\caption{ The distribution of integrated flux densities for the CGPS sources
(solid histogram) and for the NVSS catalog over the CGPS survey area (dashed
line).  \label{fig_fluxdist}
}  
\end{center}
\end{figure}

Our survey is comparable in resolution and sensitivity to the NRAO VLA Sky
Survey \citep{1998AJ....115.1693C}; the flux density scale and source positions
of the CGPS at 1420~MHz were tied to the NVSS. Figure~\ref{fig_fluxdist} 
shows
the distribution of source flux densities for the CGPS and for the NVSS over the
region of the CGPS survey. These curves give an indication of the relative
depths of the catalogs.  The NVSS flux distribution peaks at 3 mJy and drops to
zero by 2 mJy. The CGPS source  numbers continue to increase to 2 mJy, with 20\%
of the sources having flux density below 2 mJy.  This is consistent with the
relative noise levels of $\sim$0.45 mJy for the NVSS and $\sim$0.24 mJy for the
CGPS. Moreover, the CGPS, with an almost filled $u-v$  plane coverage, has a
much more uniform image background noise distribution, with very  little 
contamination from sidelobes of the synthesized beam.
Given the strong and highly structured diffuse emission that is
ubiquitous in the plane of the Galaxy, the depth of the survey will vary 
significantly with location in the plane, but should be complete to
about 2 mJy in quiet regions.

\begin{figure}[tbh]
\begin{center}
\includegraphics[width=1.05\columnwidth]{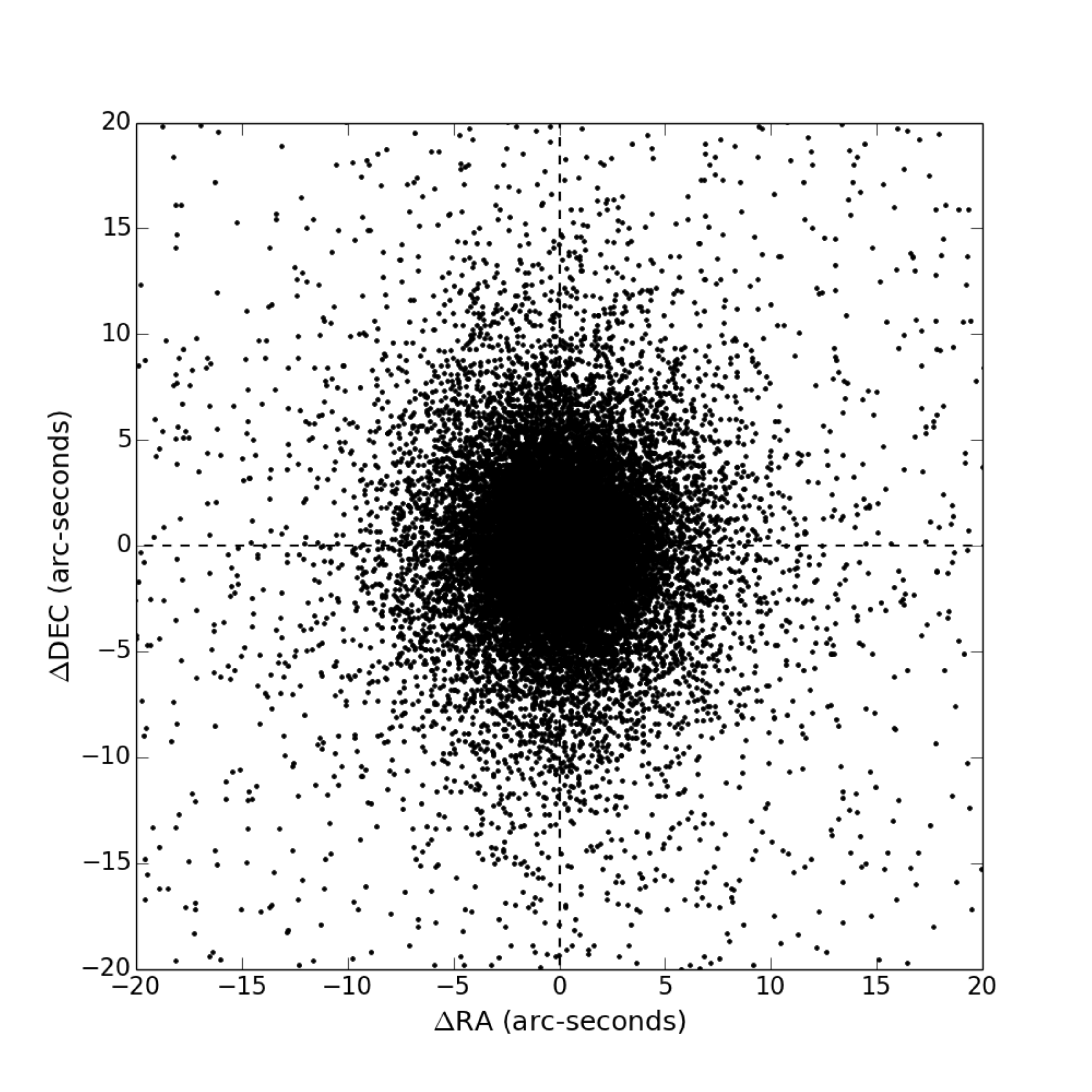}
\caption{ The differences in right ascension and declination between
the CGPS and NVSS catalog positions.  
\label{fig_positions}
}  
\end{center}
\end{figure}

From the 72,787 sources in the CGPS catalog we identified 52,876 that have
counterparts in the NVSS within 60$''$ of the CGPS position.  Figure
\ref{fig_positions} shows a plot of the position differences between the CGPS
and NVSS  catalog sources for 34,449 sources with CGPS flux density above 5
mJy.    The individual CGPS images are registered in position and flux density
against the NVSS images using a small number of strong sources within each field 
\citep{2003AJ....125.3145T}.  
Using the very large number of sources in the catalog comparison we measure an
inverse signal-to-noise weighted mean difference between the two catalogs of
$\Delta\alpha =0.011 \pm 0.003''$ and $\Delta\delta = -0.024 \pm 0.005''$.  The
difference is small, but significant  relative to the error.   To make the CGPS
catalog positions formally consistent on average with the NVSS to within
0.005$''$,  we have corrected for this small difference for the source
coordinates listed in Table \ref{catalog}.

\begin{figure}[tbh]
\begin{center}
\includegraphics[width=1.1\columnwidth]{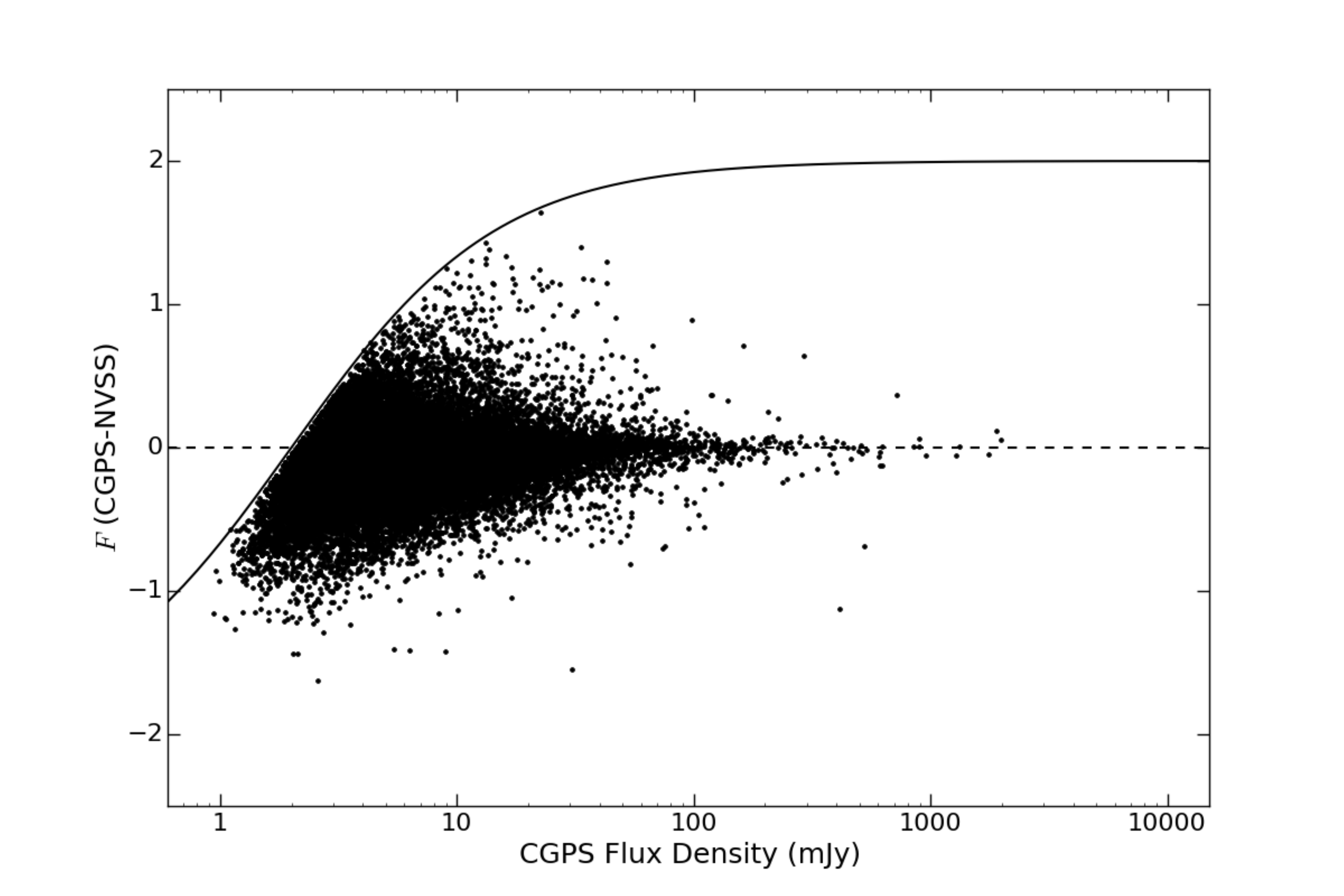}
\caption{ The fractional difference in flux density between the CGPS and NVSS
for all CGPS sources with a  counterpart in the NVSS catalog within 10$''$. The
solid line shows the effect of a detection threshold in the  NVSS flux densities of 2 mJy.
\label{fig_fluxscatter}
}  
\end{center}
\end{figure}

 To compare the CGPS and NVSS flux densities we calculate for all sources
 detected in  both surveys the fractional difference in flux density,  $F$, 
 defined as
  \begin{equation}
 F = 2  \Bigl ( \frac{S_{\rm CGPS} - S_{\rm NVSS}}{S_{\rm CGPS} + S_{\rm NVSS}} \Bigr ).
 \label{eq_F}
 \end{equation}
The result is plotted in Figure \ref{fig_fluxscatter}  for sources detected with
a position difference between the two surveys of less than 10$''$, and CGPS
major-axis diameter less than 20$''$. The solid line in the figure, that delimits the upper
bound of the data points, shows the selection effect of a 2 mJy detection threshold in
the NVSS, which is approximately the NVSS minimum detectable flux density.  
The absence of points above this line is due to noise variations in
the NVSS data reducing the NVSS signals of CGPS sources below the NVSS detection
limit.  We used 370 unresolved sources with $S_{\rm CGPS} > 40$ mJy and
position differences less than 10$''$ to derive a median fractional difference
of  $< F >$ = -0.0289$\pm$ 0.005. The raw CGPS flux densities are in the
median 2.9\% lower than the NVSS catalog flux densities. We have corrected for
this in the CGPS catalog.  The CGPS and NVSS flux density scales are thus
aligned with an error of 0.5\%.

\begin{figure}[tbh]
\begin{center}
\includegraphics[width=1.1\columnwidth]{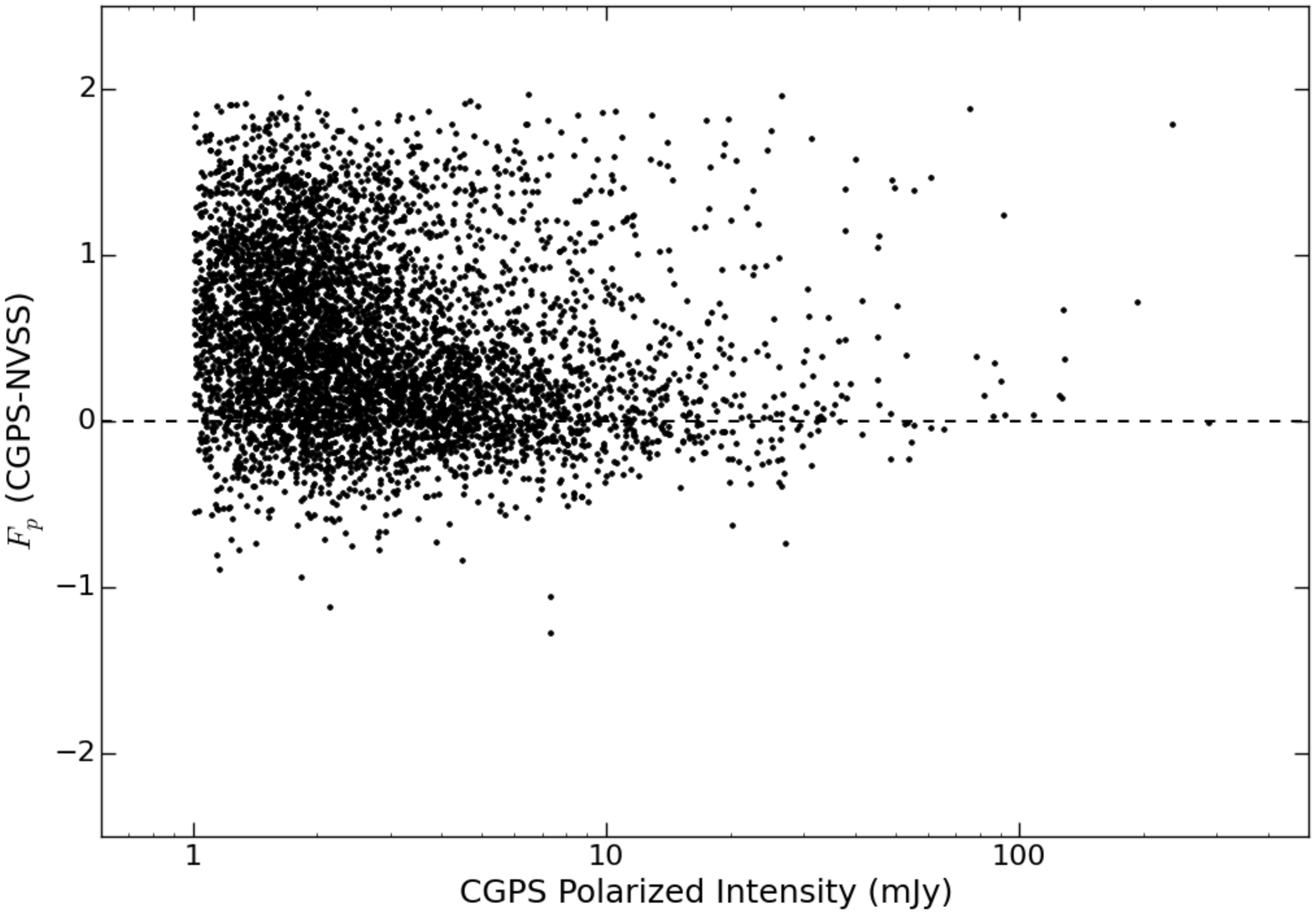}
\caption{ The fractional difference in peak polarized intensity between CGPS and NVSS for all sources 
detected in polarization in 
the CGPS having NVSS counterparts within 10$''$. 
\label{fig_pol} }   
\end{center} 
\end{figure}

\subsection{Polarization}
\label{poln}

Figure~\ref{fig_pol} shows the fractional difference in peak polarized flux
density, $F_p$, between the CGPS  and NVSS for polarized sources in the CGPS.
The data show that for the CGPS polarized intensities are on average
substantially larger than the NVSS.  This arises from the effect on the NVSS
polarized intensities of the high levels of Faraday Rotation  from propagation
through the magneto-ionic medium of the Galaxy at the low Galactic latitudes of
the CGPS.  

\begin{figure}
\begin{center}
\includegraphics[width=1.06\columnwidth]{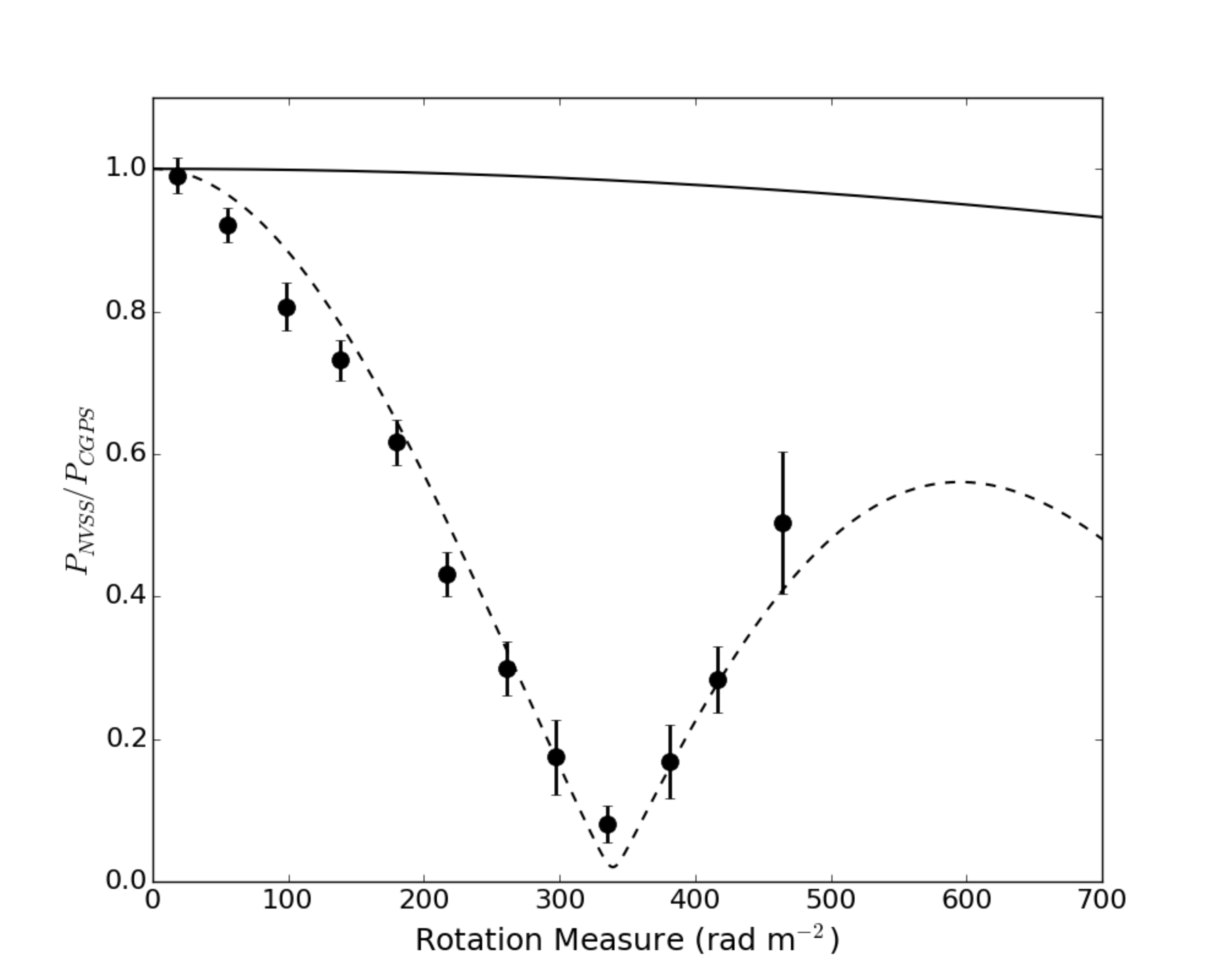}
\caption{
Comparison of NVSS and CGPS polarized intensity as a function of
Rotation Measure. The dashed line shows the expected fractional bandwidth
depolarization for the NVSS band structure.   The solid line shows the same
quantity for the CGPS.  The data points are average values of the ratio of NVSS
to CGPS polarized intensity  as a function of source Rotation Measure. The NVSS
bandwidth depolarization accounts for the systematically higher CGPS
polarizations shown in Figure~\ref{fig_pol}.
 \label{fig_depol}
}  
\end{center}
\end{figure}

As noted by \cite{1998AJ....115.1693C}, the widely separated dual-band
structure of the NVSS survey produced significant depolarization in the
band-average polarized intensities for Rotation Measure (RM) magnitudes larger than about 100
rad\,m$^{-1}$.  Over the area of the CGPS, RM values of this order and larger are
common \citep{2003ApJS..145..213B}.  The depolarization effect is illustrated by
the data points in Figure~\ref{fig_depol},  which show, as a function of source RM, 
the mean ratio of NVSS to
CGPS peak polarized intensity for CGPS  polarized sources having $p_{\rm CGPS} >
10$ mJy.  Rotation Measures were derived from the two-bands of the NVSS data
\citep{2009ApJ...702.1230T}.  The ratio recovers the theoretical depolarization
curve of the NVSS band structure  (shown by the dashed line in the Figure). The
narrower band CGPS data suffer less than 10\% depolarization up to Rotation
Meaures of $\sim$1000 rad\,m$^{-2}$.   The CGPS  polarization data thus provide
a low-latitude complement to the NVSS polarization catalog. 

\subsection{Variable and Transient Sources}
\label{vari}
 
The CGPS observations were taken over the period 1995 to 2009, while the NVSS
data were taken between 1993 and 1996.   The similar resolutions and
sensitivities of the two surveys allow for the identification of sources in the
plane of the Galaxy that have variable flux density over time intervals of a few to
several years to sensitivities of a few mJy by comparison of the catalog data.

Several candidate variable sources are visible in Figure~\ref{fig_fluxscatter}
as outliers to the noise-broadened  distribution of $F$.  However, some of these
are spurious, resulting from differences between the NVSS and CGPS algorithms
for identifying and fitting sources as either close double or extended.  In some cases 
objects fit as single extended sources in the NVSS are fit as multiples in the CGPS, and
vice-versa. Extended single component sources that are split into multiple
components in the other catalog, will have very different flux densities, and
each component will be offset in position from the single fit. 
To remove this effect from the variability analysis,
we have restricted the search for variability to unresolved CGPS sources that
are isolated from their nearest CGPS neighbour by at least 90$''$, and whose
NVSS counterpart {\bf  also classed as unresolved} and agrees in position within 5$''$. 
We also restrict the analysis to sources with
$S_{\rm CGPS} > 4.0$\,mJy. There are 10,897 CGPS sources that meet these criteria.

\begin{figure}[tbh]
\begin{center}
\includegraphics[width=0.96\columnwidth]{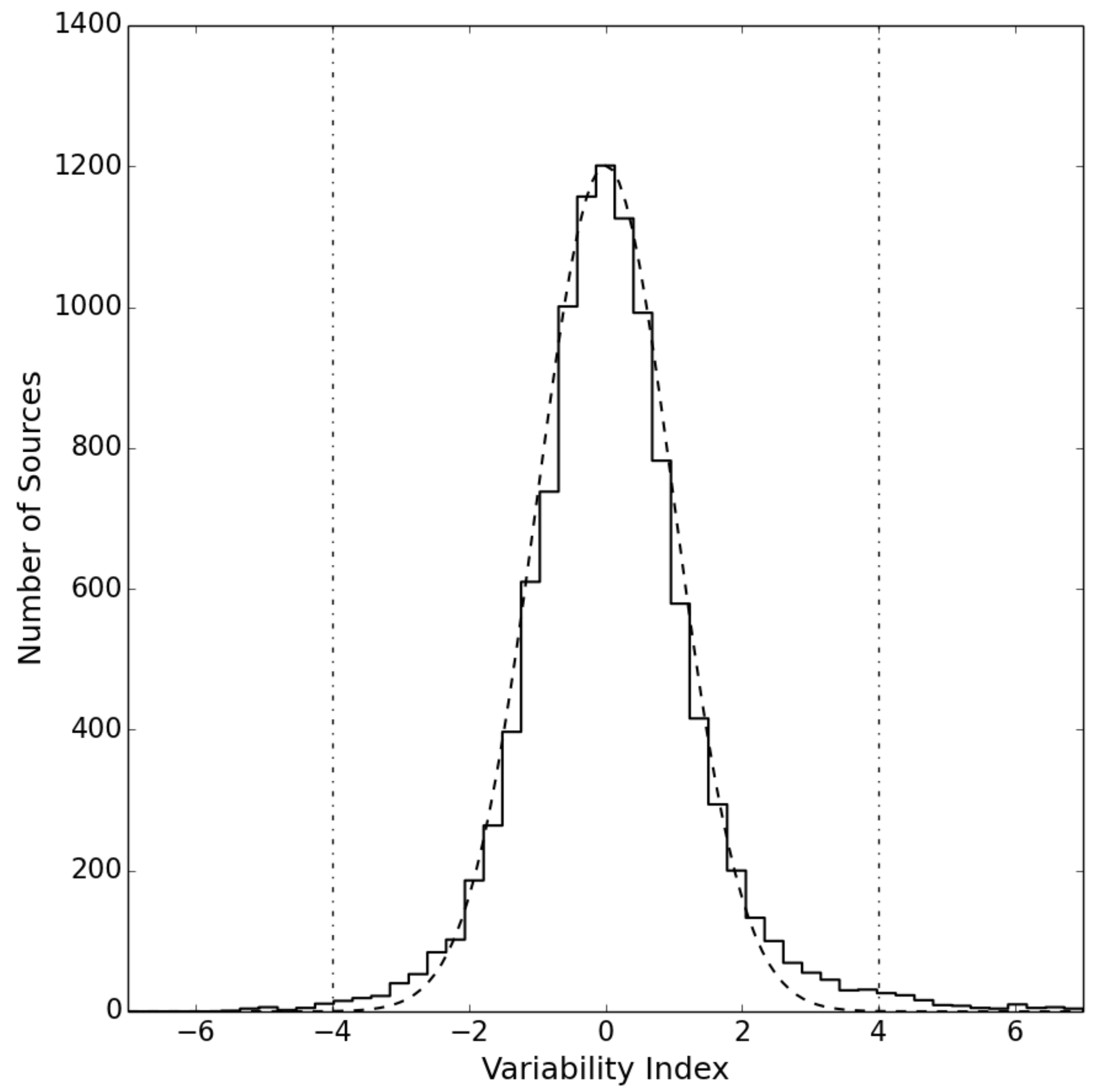}
\caption{
Distribution of variability indices, $V$.  The solid histogram is for sources
with CGPS flux density greater than 4 mJy.  The dashed curve is a Gaussian
distribution with standard deviation of unity normalised to the same peak as the
histogram.   The vertical dot-dashed lines indicate the limit of $V$ for
classification as a variable.
 \label{fig_V}
}  
\end{center}
\end{figure}

As shown in Equation~\ref{eq_F}, the quantity $F$ measures the change in flux
density between  CGPS and NVSS as a fraction of the mean flux density measured
between the two surveys.  A change in flux density by a factor of 2 results in
F = $\pm$0.67.  The significance of the change is quantified by a
comparison to the uncertainty on $F$.  We calculate this uncertainty,
$\sigma_F$, by propagating the errors on the individual flux densities $S_{\rm
CGPS}$ and $S_{\rm NVSS}$.   We then identify as variables sources those for which 
\begin{equation}
V=  \frac{F -F_{\rm MED}}{\sigma_{F}} > 4
\label{eq_V}
\end{equation}
where $F_{\rm MED}$ is the median value of $F$ for population.
In the absence of variable sources, if $\sigma_F$ is an accurate measure of the error, 
the variance in $F$ will be determined by noise
alone and the distribution of $V$ over the ensemble of sources will be Gaussian with unit standard
deviation.  Figure ~\ref{fig_V} shows the distribution of $V$ for the CGPS
variable search sample. The central peak of the distribution is well fit by the
Gaussian.  Evidence for a population of variable sources appears as excess over
the Gaussian distribution  in the wings,  beginning  at  about $|V| \sim 2$.
However, we take a conservative approach and class as a variable sources those for which
$|V| > 4$.   Of the 10,897 sources examined, 146 (1.3\%) satisfy this
criterion.   For a Gaussian distribution we expect less than 1 object at $|V| >
4$ by chance.

Figure~\ref{fig_V} shows an asymmetry in the tail of the distribution, with 
more objects with high positive $V$.   There are
111 sources with $V>4$, compared to 35 with $V<-4$.  This is a result of
the threshold in $S_{\rm CGPS}$, which results in a bias against
large negative values of $F$. A source with $S_{\rm CGPS}> 4$\,mJy and $F < -1.0$
must have $S_{\rm NVSS}> 12$\,mJy, which represents a small
fraction of the NVSS sources.  Conversely,  $F > +1.0$ requires 
$S_{\rm NVSS}> 1.25$\,mJy, sampling the entire NVSS catalog.
 
While not a complete statistical study for sources of variable emission in the
Galactic plane at 1420 MHz,  this simple comparison does identify a small sample of isolated,
unresolved  radio sources whose flux density exhibits a  large fractional change
between the two catalogs.  The resulting list of variable source candidates is
given in Table~\ref{variables}.

\begin{deluxetable}{crrrr}
\tabletypesize{\scriptsize}
\tablewidth{0pt}
\tablecaption{Variable Sources Detected in CGPS and NVSS \label{variables} }
\tablehead{
\colhead{Name}&  \colhead{S$_{\rm CGPS}$} & \colhead{S$_{\rm NVSS}$} &
\colhead{F} & \colhead{V} \\
& \colhead{(mJy)} &  \colhead{(mJy)}  
}
\startdata
CGPS J000636+673019 &     14.8 $\pm$    0.6 &      6.9 $\pm$    0.5 &    0.73 &    5.38 \\ 
CGPS J000728+671954 &     38.2 $\pm$    1.7 &      9.7 $\pm$    0.5 &    1.19 &    8.48 \\ 
CGPS J001652+642112 &     12.1 $\pm$    0.5 &      6.8 $\pm$    0.5 &    0.56 &    4.24 \\ 
CGPS J002359+644044 &     16.0 $\pm$    0.6 &      8.7 $\pm$    0.5 &    0.59 &    5.27 \\ 
CGPS J002606+644250 &      9.5 $\pm$    0.4 &      2.8 $\pm$    0.4 &    1.09 &    5.94 \\ 
CGPS J002836+631124 &     13.1 $\pm$    0.5 &      4.7 $\pm$    0.5 &    0.95 &    6.43 \\ 
CGPS J011455+605142 &     13.6 $\pm$    0.5 &      7.4 $\pm$    0.5 &    0.59 &    4.56 \\ 
CGPS J013309+610831 &      9.8 $\pm$    0.4 &     17.9 $\pm$    0.7 &   -0.58 &   -4.96 \\ 
CGPS J020039+650454 &     13.4 $\pm$    0.5 &      5.0 $\pm$    0.5 &    0.91 &    6.56 \\ 
CGPS J020222+652747 &      9.9 $\pm$    0.4 &      3.3 $\pm$    0.4 &    1.00 &    6.07 \\ 
CGPS J021722+615647 &     34.9 $\pm$    1.1 &     17.5 $\pm$    0.7 &    0.66 &    7.04 \\ 
CGPS J022303+612945 &     18.1 $\pm$    0.8 &      4.8 $\pm$    0.5 &    1.16 &    7.59 \\ 
CGPS J025804+603843 &     11.2 $\pm$    0.5 &      3.9 $\pm$    0.4 &    0.96 &    6.14 \\ 
CGPS J032425+552044 &     14.7 $\pm$    0.5 &      7.8 $\pm$    0.5 &    0.62 &    5.05 \\ 
CGPS J032556+554433 &     13.7 $\pm$    0.5 &      5.9 $\pm$    0.5 &    0.79 &    5.62 \\ 
CGPS J032641+543419 &     16.8 $\pm$    0.6 &     10.8 $\pm$    0.5 &    0.43 &    4.23 \\ 
CGPS J032853+544254 &      4.5 $\pm$    0.4 &     10.2 $\pm$    0.5 &   -0.78 &   -4.48 \\ 
CGPS J032924+531748 &     12.5 $\pm$    0.5 &      6.6 $\pm$    0.5 &    0.62 &    4.32 \\ 
CGPS J035853+541313 &     20.4 $\pm$    1.0 &     10.0 $\pm$    0.5 &    0.68 &    5.09 \\ 
CGPS J040012+490824 &      4.9 $\pm$    0.3 &      9.8 $\pm$    0.5 &   -0.66 &   -4.03 \\ 
CGPS J040509+510408 &     13.5 $\pm$    0.5 &      7.3 $\pm$    0.5 &    0.59 &    4.82 \\ 
CGPS J040916+511511 &     11.6 $\pm$    0.5 &      3.2 $\pm$    0.5 &    1.13 &    6.39 \\ 
CGPS J041125+501320 &      6.7 $\pm$    0.3 &      2.4 $\pm$    0.4 &    0.95 &    4.55 \\ 
CGPS J041329+514525 &      7.5 $\pm$    0.3 &      3.3 $\pm$    0.4 &    0.78 &    4.26 \\ 
CGPS J043122+444202 &     12.8 $\pm$    0.4 &      7.3 $\pm$    0.5 &    0.54 &    4.45 \\ 
CGPS J044709+524525 &      7.6 $\pm$    0.5 &      2.5 $\pm$    0.4 &    1.01 &    4.30 \\ 
CGPS J045000+524159 &      7.8 $\pm$    0.5 &     15.3 $\pm$    0.6 &   -0.65 &   -4.95 \\ 
CGPS J045000+452234 &      7.7 $\pm$    0.4 &      2.2 $\pm$    0.4 &    1.11 &    5.41 \\ 
CGPS J045018+484937 &     10.5 $\pm$    0.4 &      5.9 $\pm$    0.4 &    0.56 &    4.48 \\ 
CGPS J045246+430819 &     14.7 $\pm$    0.5 &     24.8 $\pm$    0.8 &   -0.51 &   -5.30 \\ 
CGPS J045808+441958 &     34.6 $\pm$    1.1 &     22.0 $\pm$    0.8 &    0.45 &    5.20 \\ 
CGPS J045849+472809 &      7.8 $\pm$    0.3 &     13.7 $\pm$    0.6 &   -0.54 &   -4.20 \\ 
CGPS J050544+360532 &      6.1 $\pm$    0.4 &     11.2 $\pm$    0.5 &   -0.59 &   -4.12 \\ 
CGPS J050749+411450 &     16.3 $\pm$    0.6 &     10.0 $\pm$    0.5 &    0.48 &    4.47 \\ 
CGPS J051704+442228 &     16.9 $\pm$    0.6 &     25.4 $\pm$    0.9 &   -0.40 &   -3.96 \\ 
CGPS J052335+405711 &      7.0 $\pm$    0.3 &     12.8 $\pm$    0.6 &   -0.59 &   -4.26 \\ 
CGPS J052902+295502 &      6.4 $\pm$    0.3 &     12.9 $\pm$    0.6 &   -0.67 &   -4.83 \\ 
CGPS J053751+351648 &      7.0 $\pm$    0.4 &     13.9 $\pm$    0.6 &   -0.66 &   -4.93 \\ 
CGPS J054126+240405 &     24.7 $\pm$    0.8 &     13.2 $\pm$    0.6 &    0.61 &    6.16 \\ 
CGPS J054519+260810 &     12.3 $\pm$    0.5 &      5.0 $\pm$    0.4 &    0.84 &    6.34 \\ 
CGPS J054604+344009 &     25.2 $\pm$    0.8 &     38.6 $\pm$    1.2 &   -0.42 &   -4.61 \\ 
CGPS J054726+225944 &     15.1 $\pm$    0.5 &      7.3 $\pm$    0.5 &    0.70 &    5.91 \\ 
CGPS J055034+330022 &      4.4 $\pm$    0.3 &      8.9 $\pm$    0.5 &   -0.68 &   -3.89 \\ 
CGPS J055144+203124 &     12.2 $\pm$    0.6 &     21.8 $\pm$    0.8 &   -0.56 &   -4.85 \\ 
CGPS J055155+235051 &     56.5 $\pm$    1.7 &     38.3 $\pm$    1.2 &    0.38 &    4.82 \\ 
CGPS J055206+200719 &     61.4 $\pm$    1.8 &     41.3 $\pm$    1.3 &    0.39 &    4.89 \\ 
CGPS J055219+201313 &      9.3 $\pm$    0.4 &     16.0 $\pm$    0.6 &   -0.53 &   -4.63 \\ 
CGPS J055327+223112 &     34.5 $\pm$    1.1 &     22.8 $\pm$    0.8 &    0.41 &    4.63 \\ 
CGPS J055532+255852 &     34.8 $\pm$    1.1 &     23.3 $\pm$    0.8 &    0.40 &    4.64 \\ 
CGPS J055951+265703 &     10.4 $\pm$    0.4 &     16.0 $\pm$    0.6 &   -0.43 &   -3.81 \\ 
CGPS J060044+254112 &     13.6 $\pm$    0.5 &      6.3 $\pm$    0.5 &    0.73 &    5.65 \\ 
CGPS J060046+193941 &     45.9 $\pm$    1.4 &     22.8 $\pm$    0.8 &    0.67 &    7.61 \\ 
CGPS J060245+301740 &      6.5 $\pm$    0.5 &     14.2 $\pm$    0.6 &   -0.74 &   -5.16 \\ 
CGPS J060329+194946 &      4.6 $\pm$    0.3 &      9.3 $\pm$    0.5 &   -0.68 &   -3.90 \\ 
CGPS J060542+314556 &      4.5 $\pm$    0.5 &      9.9 $\pm$    0.5 &   -0.75 &   -4.04 \\ 
CGPS J060547+311443 &      5.2 $\pm$    0.4 &     11.0 $\pm$    0.5 &   -0.72 &   -4.46 \\ 
CGPS J060821+282942 &      9.6 $\pm$    0.4 &     17.6 $\pm$    0.7 &   -0.58 &   -5.02 \\ 
CGPS J060838+195646 &     19.2 $\pm$    0.7 &      9.7 $\pm$    0.5 &    0.66 &    6.09 \\ 
CGPS J060900+201315 &     35.1 $\pm$    1.1 &     22.6 $\pm$    0.8 &    0.43 &    4.99 \\ 
CGPS J061305+200914 &     37.9 $\pm$    1.2 &     22.3 $\pm$    0.8 &    0.52 &    5.84 \\ 
CGPS J061420+183700 &      7.7 $\pm$    0.5 &     14.0 $\pm$    0.6 &   -0.58 &   -4.10 \\ 
CGPS J061659+230942 &     25.9 $\pm$    0.8 &      6.7 $\pm$    0.5 &    1.18 &   10.45 \\ 
CGPS J061702+233446 &      7.0 $\pm$    0.3 &      2.6 $\pm$    0.4 &    0.92 &    4.48 \\ 
CGPS J062652+213422 &     19.9 $\pm$    0.7 &     12.8 $\pm$    0.6 &    0.43 &    4.21 \\ 
CGPS J192307+223405 &      5.0 $\pm$    0.3 &      9.7 $\pm$    0.5 &   -0.64 &   -4.01 \\ 
CGPS J192409+252948 &      5.5 $\pm$    0.3 &     11.6 $\pm$    0.5 &   -0.72 &   -5.30 \\ 
CGPS J193124+224331 &    736.6 $\pm$   21.5 &    495.0 $\pm$   14.9 &    0.39 &    5.04 \\ 
CGPS J193452+201417 &      4.6 $\pm$    0.3 &      9.1 $\pm$    0.5 &   -0.65 &   -3.89 \\ 
CGPS J193632+220847 &     18.8 $\pm$    0.7 &     11.6 $\pm$    0.5 &    0.47 &    4.45 \\ 
CGPS J193633+265436 &     19.2 $\pm$    0.7 &      8.5 $\pm$    0.5 &    0.77 &    6.77 \\ 
CGPS J193648+205135 &     58.6 $\pm$    1.7 &     30.2 $\pm$    1.0 &    0.64 &    7.57 \\ 
CGPS J193707+311520 &      6.7 $\pm$    0.4 &     11.7 $\pm$    0.5 &   -0.55 &   -4.03 \\ 
CGPS J193939+271240 &     17.5 $\pm$    0.6 &     11.4 $\pm$    0.5 &    0.42 &    4.39 \\ 
CGPS J194048+271050 &      8.8 $\pm$    0.4 &      3.9 $\pm$    0.4 &    0.78 &    4.89 \\ 
CGPS J194201+171141 &     24.0 $\pm$    0.8 &     40.0 $\pm$    1.3 &   -0.50 &   -5.27 \\ 
CGPS J194537+325738 &     23.6 $\pm$    0.7 &     16.0 $\pm$    0.6 &    0.38 &    4.34 \\ 
CGPS J194728+211157 &     64.2 $\pm$    1.9 &     37.4 $\pm$    1.2 &    0.53 &    6.46 \\ 
CGPS J194951+210656 &    207.0 $\pm$    6.4 &    138.1 $\pm$    4.2 &    0.40 &    4.89 \\ 
CGPS J195228+372724 &     11.2 $\pm$    0.5 &      5.6 $\pm$    0.5 &    0.67 &    4.43 \\ 
CGPS J195737+363830 &     12.2 $\pm$    0.4 &      3.9 $\pm$    0.5 &    1.03 &    6.70 \\ 
CGPS J195915+345843 &     17.4 $\pm$    0.6 &      7.8 $\pm$    0.5 &    0.76 &    6.72 \\ 
CGPS J200510+364336 &     11.7 $\pm$    0.4 &      6.3 $\pm$    0.5 &    0.60 &    4.45 \\ 
CGPS J200702+401729 &     58.6 $\pm$    2.8 &     32.9 $\pm$    1.2 &    0.56 &    4.61 \\ 
CGPS J200855+300842 &      7.7 $\pm$    0.3 &     13.6 $\pm$    0.6 &   -0.55 &   -4.30 \\ 
CGPS J200915+403931 &     68.9 $\pm$    2.8 &     31.8 $\pm$    1.1 &    0.74 &    6.63 \\ 
CGPS J200921+372739 &     15.8 $\pm$    0.5 &      9.1 $\pm$    0.6 &    0.54 &    4.54 \\ 
CGPS J200925+375937 &     29.8 $\pm$    0.9 &     16.2 $\pm$    0.7 &    0.59 &    6.31 \\ 
CGPS J200931+372420 &      9.0 $\pm$    0.4 &      3.9 $\pm$    0.5 &    0.79 &    4.33 \\ 
CGPS J201008+395741 &     23.5 $\pm$    1.2 &      6.6 $\pm$    0.7 &    1.12 &    6.49 \\ 
CGPS J201148+391855 &     21.0 $\pm$    0.8 &      9.3 $\pm$    0.7 &    0.77 &    5.99 \\ 
CGPS J201528+412358 &     29.0 $\pm$    1.0 &     13.2 $\pm$    0.7 &    0.75 &    6.73 \\ 
CGPS J202036+405756 &    165.9 $\pm$    5.2 &     76.8 $\pm$    2.4 &    0.73 &    8.29 \\ 
CGPS J202241+435510 &     29.0 $\pm$    1.0 &     13.5 $\pm$    0.7 &    0.73 &    6.78 \\ 
CGPS J202450+370258 &     23.6 $\pm$    0.8 &     14.0 $\pm$    0.6 &    0.51 &    5.13 \\ 
CGPS J202516+370920 &     11.6 $\pm$    0.5 &      4.3 $\pm$    0.5 &    0.92 &    5.74 \\ 
CGPS J202527+424202 &     14.7 $\pm$    0.7 &      3.9 $\pm$    0.5 &    1.16 &    6.37 \\ 
CGPS J202726+380538 &     13.4 $\pm$    0.5 &      5.6 $\pm$    0.6 &    0.82 &    5.18 \\ 
CGPS J202751+384325 &     12.7 $\pm$    0.5 &      5.9 $\pm$    0.6 &    0.73 &    4.55 \\ 
CGPS J202824+403750 &     39.8 $\pm$    1.3 &     12.8 $\pm$    0.7 &    1.03 &    9.44 \\ 
CGPS J202848+324618 &     17.1 $\pm$    0.8 &     28.2 $\pm$    0.9 &   -0.49 &   -4.46 \\ 
CGPS J202924+401114 &     51.6 $\pm$    1.7 &     26.1 $\pm$    1.0 &    0.66 &    6.89 \\ 
CGPS J203012+381204 &     11.9 $\pm$    0.4 &      3.6 $\pm$    0.7 &    1.07 &    5.33 \\ 
CGPS J203217+372407 &     18.8 $\pm$    0.6 &     10.5 $\pm$    0.6 &    0.56 &    5.27 \\ 
CGPS J203227+381010 &     24.6 $\pm$    0.8 &     15.2 $\pm$    0.7 &    0.47 &    4.80 \\ 
CGPS J203424+453155 &     14.0 $\pm$    0.5 &      2.5 $\pm$    0.5 &    1.39 &    8.65 \\ 
CGPS J203445+363545 &      9.1 $\pm$    0.3 &      4.4 $\pm$    0.5 &    0.70 &    4.14 \\ 
CGPS J203827+431329 &     17.6 $\pm$    0.7 &      4.4 $\pm$    0.6 &    1.20 &    7.89 \\ 
CGPS J203916+430155 &     26.3 $\pm$    0.9 &     17.0 $\pm$    0.7 &    0.43 &    4.51 \\ 
CGPS J204230+404520 &      8.3 $\pm$    0.4 &      2.8 $\pm$    0.6 &    0.99 &    4.15 \\ 
CGPS J204258+373537 &      7.0 $\pm$    0.3 &      2.7 $\pm$    0.4 &    0.88 &    4.25 \\ 
CGPS J204321+390628 &     21.2 $\pm$    0.7 &     12.5 $\pm$    0.7 &    0.52 &    4.66 \\ 
CGPS J204951+424745 &     13.6 $\pm$    0.6 &      2.7 $\pm$    0.5 &    1.34 &    7.36 \\ 
CGPS J204955+642258 &     17.2 $\pm$    0.7 &     26.0 $\pm$    0.9 &   -0.41 &   -3.83 \\ 
CGPS J205021+510906 &     15.0 $\pm$    0.6 &      9.0 $\pm$    0.5 &    0.50 &    4.38 \\ 
CGPS J205032+390719 &     45.8 $\pm$    1.4 &     75.7 $\pm$    2.3 &   -0.49 &   -5.52 \\ 
CGPS J205205+431013 &     39.2 $\pm$    1.2 &     57.4 $\pm$    1.8 &   -0.38 &   -4.10 \\ 
CGPS J205408+414321 &     25.7 $\pm$    0.8 &     37.8 $\pm$    1.2 &   -0.38 &   -4.07 \\ 
CGPS J205800+421048 &     24.2 $\pm$    0.8 &     14.0 $\pm$    0.6 &    0.53 &    5.56 \\ 
CGPS J210511+493405 &     10.2 $\pm$    0.4 &      5.2 $\pm$    0.5 &    0.65 &    4.25 \\ 
CGPS J210802+522105 &      8.3 $\pm$    0.3 &      3.5 $\pm$    0.5 &    0.81 &    4.19 \\ 
CGPS J210821+502815 &     21.2 $\pm$    0.8 &     13.1 $\pm$    0.6 &    0.47 &    4.58 \\ 
CGPS J210905+522627 &     14.7 $\pm$    0.6 &      8.4 $\pm$    0.5 &    0.55 &    4.45 \\ 
CGPS J211955+492245 &     11.7 $\pm$    0.4 &      6.3 $\pm$    0.5 &    0.60 &    4.36 \\ 
CGPS J212331+522844 &     14.1 $\pm$    0.6 &      7.6 $\pm$    0.5 &    0.60 &    4.54 \\ 
CGPS J212511+574220 &     23.5 $\pm$    0.9 &     36.4 $\pm$    1.2 &   -0.43 &   -4.23 \\ 
CGPS J213835+500901 &     18.9 $\pm$    0.6 &     10.0 $\pm$    0.6 &    0.62 &    5.60 \\ 
CGPS J214148+545858 &     12.8 $\pm$    0.5 &      6.8 $\pm$    0.5 &    0.61 &    4.68 \\ 
CGPS J215403+531840 &      4.3 $\pm$    0.3 &      9.1 $\pm$    0.5 &   -0.71 &   -4.03 \\ 
CGPS J221021+525433 &      8.4 $\pm$    0.4 &      3.8 $\pm$    0.4 &    0.76 &    4.34 \\ 
CGPS J224530+701557 &     10.9 $\pm$    0.5 &      6.0 $\pm$    0.4 &    0.58 &    4.20 \\ 
CGPS J224645+580306 &     32.7 $\pm$    1.1 &     21.2 $\pm$    0.8 &    0.43 &    4.64 \\ 
CGPS J225359+644422 &     11.9 $\pm$    0.5 &      6.3 $\pm$    0.5 &    0.62 &    4.47 \\ 
CGPS J225522+581311 &      8.8 $\pm$    0.4 &      2.4 $\pm$    0.5 &    1.14 &    5.18 \\ 
CGPS J230114+572047 &      6.6 $\pm$    0.3 &      2.3 $\pm$    0.4 &    0.96 &    4.41 \\ 
CGPS J230948+622935 &     13.0 $\pm$    0.5 &      7.4 $\pm$    0.5 &    0.55 &    4.42 \\ 
CGPS J233748+655723 &     10.4 $\pm$    0.4 &      5.0 $\pm$    0.5 &    0.70 &    4.36 \\ 
CGPS J234042+614508 &      9.4 $\pm$    0.4 &      3.2 $\pm$    0.6 &    0.99 &    4.65 \\ 
CGPS J234205+613806 &      8.9 $\pm$    0.4 &      3.3 $\pm$    0.5 &    0.92 &    4.79 \\ 
CGPS J234210+613508 &     14.8 $\pm$    0.5 &      5.8 $\pm$    0.6 &    0.87 &    6.13 \\ 
CGPS J234612+610229 &     15.4 $\pm$    0.5 &      7.3 $\pm$    0.5 &    0.72 &    6.08 \\ 
CGPS J235007+601852 &      9.2 $\pm$    0.4 &      2.6 $\pm$    0.4 &    1.11 &    5.96 \\ 
CGPS J235032+643035 &     12.5 $\pm$    0.5 &      6.4 $\pm$    0.5 &    0.64 &    4.70 \\ 
CGPS J235221+590043 &      7.7 $\pm$    0.4 &      2.8 $\pm$    0.4 &    0.94 &    4.86 \\ 
CGPS J235312+644017 &      9.2 $\pm$    0.4 &      3.3 $\pm$    0.4 &    0.95 &    5.69 \\ 
CGPS J235400+594207 &      7.3 $\pm$    0.3 &      3.2 $\pm$    0.4 &    0.78 &    4.19 \\ 
CGPS J235612+625132 &      7.4 $\pm$    0.4 &      3.1 $\pm$    0.4 &    0.82 &    4.26 \\ 
\enddata
\end{deluxetable}

\begin{figure*}[tbh]
\begin{center}
\includegraphics[width=\textwidth]{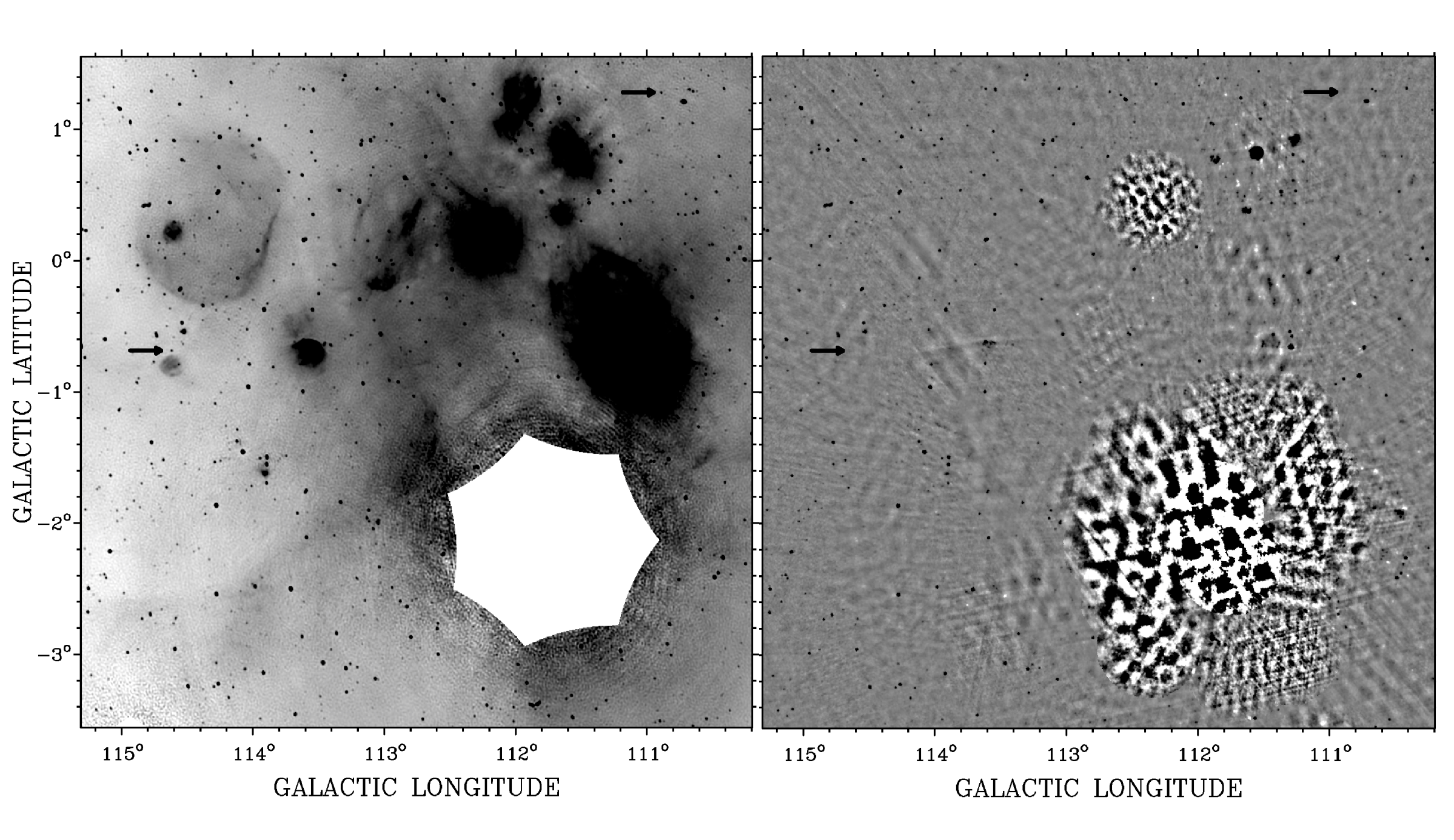}
\caption{
Comparison of CGPS (left) and NVSS (right) images of a field near Cas A. The
area around Cas A is blanked in the CGPS image. The grayscale in the CGPS images
extends from 5K (white) to 8K (black), and in the NVSS image from -0.015 (white)
to +0.015 (black) mJy/beam. Two sources that are possibly transients
are marked by arrows in the CGPS image; they cannot be detected in the 
NVSS image (see Figure~\ref{fig_transient}). 
 \label{fig_comparison}
}  
\end{center}
\end{figure*}

In addition to searching for variable sources that appear in both the CGPS and NVSS 
catalogues, we also looked for candidate transient objects that appear in the CGPS but have
no counterpart in the NVSS catalog.
There are 19,911 sources in the CGPS without NVSS counterparts.  
The vast majority of these are faint objects below $\sim$3 mJy that are 
missed by the NVSS due to the high level of NVSS imaging artefacts arising from the 
lack of low spatial frequency data in the presence of 
the strong extended emission at low Galactic latitudes. 
This effect is illustrated in Figure~\ref{fig_comparison}, where we
compare CGPS and NVSS images for one of the CGPS 5.12$^{\circ}$$\times$5.12$^{\circ}$ mosaics. 
These images include Cas A, and artefacts from this strong source are visible in both. 
The diffuse structure away from Cas A visible in the CGPS is not well represented in the NVSS image and
results in lower level artefacts in the NVSS that extend throughout the image.
The CGPS fully samples the structure of the diffuse emission resulting in much greater 
intensity dynamic range. 
The CGPS fully samples the structure of the diffuse emission resulting in much greater 
intensity dynamic range.  

\begin{figure}[tbh]
\begin{center}
\includegraphics[width=\columnwidth]{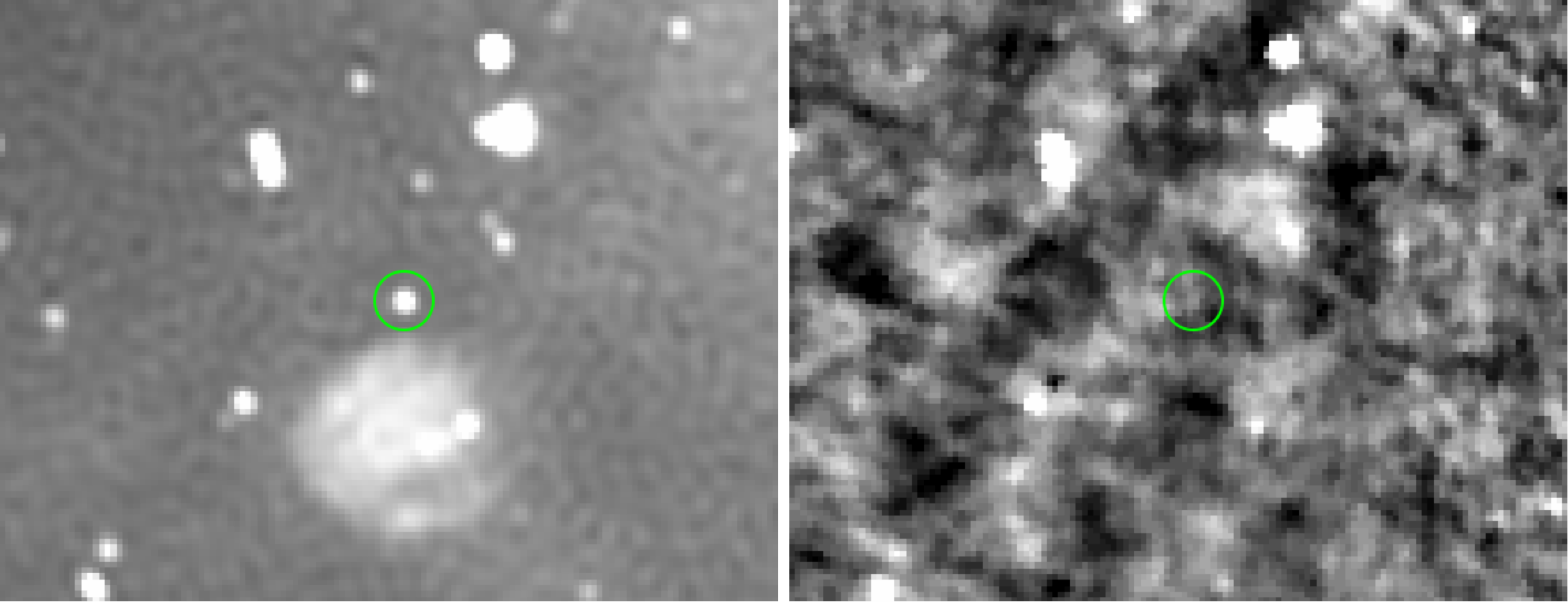}
\caption{
A close up of CGPS (left) and NVSS (right) of the area of the transient candidate
CGPS\,J234153+610329. The object indicated by the circle has flux density of 12.6 mJy 
in the CGPS catalogue.  There is no evidence of the source in the NVSS image,
which has an RMS in the location of the source of 0.58 mJy\,beam$^{-1}$.
 \label{fig_transient}
}  
\end{center}
\end{figure}

To avoid missing sources due to the effect described above, we limited the analysis for candidate transients 
to unresolved sources stronger than 10\,mJy in the CGPS.  The resulting 71 objects were each examined in the
NVSS images.  In most cases, the missing detections in the NVSS could be attributed to strong
image artefacts obscuring the source.  
 The final list of 13 transient candidates includes sources
that should have been visible in the NVSS if present at the CGPS flux levels.
These sources are listed in Table~\ref{tab_transients}.
Figure~\ref{fig_transient} shows an example of one transient candidate, CGPS\,J234153+610329, that is
detected at 12.6 mJy in the CGPS but is not visible in the NVSS down to the NVSS image RMS 
at the source location of 0.58 mJy.

\begin{deluxetable}{crr}
\tabletypesize{\scriptsize}
\tablewidth{0pt}
\tablecaption{Strong CGPS sources not detected in the NVSS \label{tab_transients} }
\tablehead{
\colhead{Name}&  \colhead{S$_{\rm CGPS}$} & \colhead{Comment}   \\
& \colhead{(mJy)} 
}
\startdata
CGPS J200258+333748 & 18.5 $\pm$ 0.6 & weak NVSS source visible   \\ 
CGPS J201912+420531  & 24.8 $\pm$ 0.9 &  \\
CGPS J201936+405852  & 20.2 $\pm$ 1.5 & \\
CGPS J202626+394015  & 17.8 $\pm$ 0.8 & \\
CGPS J202635+421335   & 18.7 $\pm$ 0.7 & weak  NVSS source visible \\
CGPS J202701+354228  & 10.0 $\pm$ 0.5 & \\
CGPS J203245+384529  &  16.2 $\pm$ 0.6 & weak NVSS source visible \\
CGPS J203315+463424  &  12.5 $\pm$ 0.6 & weak NVSS source visible \\
CGPS J221640+561817  &  10.4 $\pm$ 0.4 & weak NVSS source visible \\
CGPS J224712+580953  &  22.7 $\pm$ 0.9 & weak NVSS source visible \\
CGPS J225759+623319  &  24.6 $\pm$ 1.4 & weak NVSS source visible \\
CGPS J230704+614118  & 13.0 $\pm$ 0.5 &  \\
CGPS J234153+610329  & 12.6 $\pm$ 0.4 & \\
\enddata
\end{deluxetable}

\section{Conclusions}

We have described a catalog of 72,787 compact sources down to a minimum peak
total intensity of $\sim$1~mJy from the Canadian Galactic Plane Survey, and have
given details of electronic access to it. The catalog presents position, total
intensity and linearly polarized intensity, polarization position angle, and
source size.  The flux density scale and the positions of the CGPS were tied to
the VLA Northern Sky Survey catalog using strong, isolated CGPS sources that have counterparts in
the NVSS.  The positions in the two catalogs are aligned to systematic error of
less than $\sim0.01''$, and flux densities to within 0.5\%.
We have detected polarization from 12,368 sources. 

The effects of bandwidth
depolarization are substantially smaller in the CGPS than in  the NVSS, and the
CGPS catalog provides a superior list of polarized intensities at low Galactic
latitudes, where the NVSS polarized intensities are significantly affected by depolarization. 
Since the CGPS and the NVSS were observed at different times, we have used the comparison 
to identify variable and possibly transient sources, and list 149 variables that
are detected in both surveys, and 13 candidate transient sources detected in the CGPS
but missing in the NVSS.

\acknowledgments
The Dominion Radio Astrophysical Observatory is operated as a national facility
by the National Research Council Canada. The Canadian Galactic Plane Survey was
supported by the Natural Sciences and Engineering Research Council. The high
quality of the CGPS data owes much to the excellent work of the DRAO staff who
made the observations and maintained the telescope, and that of the CGPS team
who reduced the data.

\facility{DRAO (synthesis telescope)}

\bibliography{drao}

\end{document}